\documentclass[twocolumn, amsmath,amssymb,aps]{revtex4-1}

\usepackage{verbatim}
\usepackage{graphicx}
\usepackage{bm} 
\usepackage{ifpdf} 
\usepackage{dcolumn}  
\usepackage{epsfig}
\usepackage{color}
\usepackage[usenames,dvipsnames]{xcolor}
\usepackage{siunitx}

\def\rme{{\mathrm{e}}}

\newcommand{\kB}{k_\mathrm{B}}
\newcommand{\kk}{K$_2$SO$_4$}
\newcommand{\kkk}{K$_3$Fe(CN)$_6$}
\newcommand{\kkkk}{K$_4$Fe(CN)$_6$}
\def\Eq{Eq.}

\newcommand{\celsius}{\ensuremath{^\circ}C}

\begin{document}

\title{Interactions between Silica Particles in the Presence of Multivalent Coions}
\author{Biljana Uzelac}
\author{Valentina Valmacco}
\email{Current Address: Firmenich SA, Corporate R\&D Division, Rue de la Bergère 7, 1217 Geneva, Switzerland}
\author{Gregor Trefalt}
\email{E-mail: \texttt{gregor.trefalt@unige.ch}}

\affiliation{Department of Inorganic and Analytical Chemistry, University of Geneva,
Sciences II, 30 Quai Ernest-Ansermet, 1205 Geneva,
Switzerland}

\date{\today}

\begin{abstract}
Forces between charged silica particles in solutions of multivalent {\sl coions} are measured with colloidal probe technique
based on atomic force microscopy. The concentration of 1:$z$ electrolytes is systematically varied to understand the
behavior of electrostatic interactions and double-layer properties in these systems. Although the coions are multivalent the 
Derjaguin, Landau, Verwey, and Overbeek (DLVO) theory perfectly describes the measured force profiles. The diffuse-layer 
potentials and regulation properties are extracted from the forces profiles by using the DLVO theory. The dependencies
of the diffuse-layer potential and regulation parameter shift to lower concentration with increasing coion valence when
plotted as a function of concentration of 1:$z$ salt. Interestingly, these profiles collapse to a master curve if plotted
as a function of monovalent counterion concentration.
\end{abstract}

\maketitle


\section{Introduction}

Interactions between charged objects in electrolyte solutions are important for many biological systems~\cite{Strey1998}, and 
in processes such as paper making~\cite{Iselau2015}, waste water treatment~\cite{Bolto2007}, 
ceramic processing~\cite{Cerbelaud2010}, ink-jet printing~\cite{Kuscer2012}, particle design~\cite{Zanini2017}, and
concrete hardening~\cite{Labbez2006}.  
Electrostatic interactions across such solutions are strongly influenced by the type of ions present 
and their concentration. One of the most important ion properties in this respect is the ionic valence. Furthermore, it 
is important whether the multivalent ions are {\sl counterions}, which are oppositely charged than the surface, or 
{\sl coions}, which carry the same charge as the surface.

The forces between charged colloidal particles or charged surfaces across aqueous solutions can now be routinely measured 
with variety of experimental techniques such as surface force apparatus (SFA)~\cite{Israelachvili2010, Espinosa-Marzal2012}, 
colloidal probe atomic force microscopy (AFM)~\cite{Ducker1991, Butt1991}, total internal reflection microscopy (TIRM)~
\cite{Prieve1999, vonGrunberg2001}, and optical tweezers~\cite{Gutsche2007, Crocker1994}. 

Direct force measurements in the presence of multivalent counterions received increased attention lately, especially in the 
view of the validity of their description within the mean-field Poisson-Boltzmann theory\cite{Pashley1984, Besteman2004, 
Zohar2006, Sinha2013, Danov2016, MontesRuiz-Cabello2014b, MoazzamiGudarzi2015, Valmacco2016, Trefalt2017a}. 
Multivalent counterions 
typically strongly influence the electrostatic forces already at very minute concentrations~
\cite{Besteman2004, Zohar2006,Sinha2013, MontesRuiz-Cabello2014b, MoazzamiGudarzi2015, Valmacco2016}. These studies showed
that mean-field PB theory can be used with confidence at larger separation distances, while at smaller separations of
few nanometers, the experimental forces deviate from the PB description. These deviations can be caused by ion-ion 
correlations and/or other short-range forces~\cite{MontesRuiz-Cabello2014b, Trefalt2017a, Zohar2006, Trulsson2006,
Kanduc2017}.

While systems containing multivalent counterions are relatively well investigated, literature is much scarcer in the case 
of multivalent coion systems. Force profiles across solutions containing multivalent counterions are usually
exponential, which is typical for double-layer forces. Interestingly, their coion counterparts invoke non-exponential 
and soft long-ranged forces, which can be well described with the PB theory~\cite{Kohonen2000, MontesRuiz-Cabello2014b}.
Extreme case of such non-exponential force profiles can be induced by like-charged polyelectrolytes, which can be modelled 
as coions with extremely high effective valence~\cite{Moazzami-Gudarzi2016c, Moazzami-Gudarzi2017}. This non-exponential 
behavior is induced by the expulsion of
the multivalent coions from the slit between two charged surfaces. At large-separation distances, both counterions and
coions enter in the slit. When distance between the surfaces is reduced, the electrostatic repulsion between the surface and 
multivalent coions is increased, which finally leads to the expulsion of the latter from the slit and only monovalent
counterions are left between the charged surfaces. These counterion-only double-layer results in a power-law dependence of
the force~\cite{Langmuir1938, Briscoe2002a, MontesRuiz-Cabello2014b}. Therefore, the forces in multivalent coion systems are 
exponential only at large
distances and transition to a power-law behavior at smaller separations. Furthermore, the influence on these type on
interactions on the aggregation of colloidal particles can have non-expected results. For example, for aggregating 
suspension critical coagulation ionic strength (CCIS) strongly decreases with increasing counterion valence~
\cite{Trefalt2017}. This behavior is known for over a century and is referred to as the classical Schulze-Hardy 
rule. Surprisingly, if one uses the multivalent coions as an aggregating agent as opposed to the counterions, an inversion
is observed as it was shown recently~\cite{Cao2015}. In multivalent coion systems the CCIS is increasing with increasing 
valence, $i.e.$ coagulating power of coions decreases with increasing valence. Due to this inversion, the phenomenon was 
given a name: {\sl inverse} Schulze-Hardy rule~\cite{Cao2015, Trefalt2016b, Trefalt2017}.

Although, some information about these forces exists, to the 
best of our knowledge there are no studies which systematically investigate the forces and properties of the double-layer in 
a large range of concentrations for different coion valences. In particular, no detailed information exists on the 
regulation of charged surfaces in these systems.

In this work we focus on interactions between silica particles across solutions containing multivalent coions. We
systematically study the influence of the concentration and valence of the coions on double-layer forces. We further
examine the properties of the electric double-layer, such as diffuse-layer potential and regulation parameter, and try
to pinpoint crucial factors which determine its behavior. Furtheremore a direct comparison of force curves at same
salt concentration, counterion concentration, and ionic strength reveals how the structure of the double-layer changes when
the valence of the coion changes.

\section{Experimental}

\subsection{Materials}

For force measurements spherical silica particles (Bangs Laboratories Inc., USA) were used. The producer
reports an average size of 5.2~$\mu$m. Before the measurements particles were heated at 1200~\celsius\ for
2 hours. During heat treatment the particles shrink for about 15~\% which yields an average diameter of 
4.4~$\mu$m as reported earlier \cite{Valmacco2016a}. The  root  mean square (RMS) roughness of 0.63~nm was
measured by AFM imaging in liquid~\cite{Valmacco2016a}. Forces were measured in aqueous solutions of KCl 
(Sigma Aldrich), \kk\ (Acros Organics), \kkk\ (Sigma Aldrich), and \kkkk\ (Sigma Aldrich).
The pH was kept at $10\pm 0.5$ with addition of 1~mM KOH (Acros Organics) and was checked before and after each measurement.
Mili-Q water (Millipore) was use throughout.

\subsection{Force Measurements}

Force measurements were carried out with colloidal probe technique in the symmetric sphere-sphere geometry
~\cite{Trefalt2017a}. The particles were first glued to the tip-less cantilevers (MikroMasch, Tallin, Estonia) which were
beforehand cleaned in air plasma (PDC-32G, Harrick, New York) for 5 min. Tiny drop of glue (Araldite 2000+)
and few silica particles were placed on a glass slide. The cantilever was mounted in the AFM head and 
manipulated to touch the glue then a silica particle was picked up and glued on the cantilever. The particles 
were separately spread on a quartz substrate (Ted Pella inc.), which was cleaned with piranha solution (3:1 mixture of 
H$_2$SO$_4$ (98~\%) and H$_2$O$_2$ (30~\%)). Both a quartz slide and a cantilever were then heated
side-by-side in an oven at 1200~\celsius\ for 2~h. The heating procedure resulted in a firm attachment of the 
particles to the substrate and the cantilever. During this process the glue is also completely removed.

All the measurements were done at room temperature $23\pm2$~\celsius\ with a closed-loop AFM (MFP-3D, Asylum 
Research) mounted on an inverted optical microscope (Olympus IX70). The quartz slides and cantilevers with 
attached probes were rinsed with water and 
ethanol, dried in air, and plasma-treated for 20 min. The quartz substrate was glued (Pattex 100\% Repair Gel)
onto the glass slide sealing the AFM cell. The AFM fluid cell was mounted and flushed thoroughly with the 
respective electrolyte solution. The particle on the cantilever was centered above one particle on the 
substrate with the precision of about 100~nm. The deflection of the cantilever was recorded for 100-200 
approach-retract cycles with the sampling rate of 5~kHz and cantilever velocities of 300~nm/s, cycling 
frequency was 0.5~Hz. The zero separation distance was assumed when the force reached a value of 10~mN/m 
for repulsive curves, and 4~mN/m for attractive curves. Cantilever deflection was converted to the force
using Hook's law, where the spring-constant of the cantilever was determined by the method described by Sader
et al.~\cite{Sader1999}. The approach part of the raw force curves obtained with the procedure described above were
averaged. The averaging of about 150 curves leads to the noise level of about 2~pN. Only such averaged force 
profiles are used in subsequent analysis. For each condition forces between 3-5 different pairs of particles
were measured.

\section{Charging of the Silica Surface}

Basic Stern model is used to model the surface charge of silica particles at different solution compositions.
For simplicity we just use 1-p$K$ model, where only one type of silanol groups can undergo deprotonation
according to the following reaction~\cite{Hiemstra1989a, Kobayashi2005, Behrens2001}
\begin{equation}
{\rm SiOH \rightleftharpoons SiO^- + H^+} \, .
\label{eq:silanol}
\end{equation}
The equilibrium between bulk protons and silanol groups is established according to the following equation,
\begin{equation}
K = \frac{{\rm [H^+]}e^{\beta e_0\psi_0}\Gamma_{\rm SiO^-}}{\Gamma_{\rm SiOH}} \, ,
\label{eq:equilibrium}
\end{equation}
where $K$ (${\rm p}K = -\log K$) is the equilibrium constant, ${\rm [H^+]}$ (${\rm pH = -\log [H^+]}$) is the 
bulk concentration of protons, $e_0$ is the elementary charge, $\beta = 1/(\kB T)$ is the inverse thermal 
energy, $\psi_0$ is the surface potential,
and $\Gamma_{\rm SiO^-}$ and $\Gamma_{\rm SiOH}$ are the surface densities of deprotonated and protonated
silanol groups, respectively. The total number of silanol groups on the surface is given by
\begin{equation}
\Gamma_0 = \Gamma_{\rm SiO^-} + \Gamma_{\rm SiOH} \, .
\label{eq:totalDensity}
\end{equation}
The surface charge density can be calculated from the number of deprotonated silanol groups as
\begin{equation}
\sigma = -e_0\Gamma_{\rm SiO^-} \, .
\label{eq:surfChDen}
\end{equation}
The potential drop over the Stern plane is determined by Stern layer capacitance, $C_{\rm S}$,
\begin{equation}
C_{\rm S} = \frac{\sigma}{\psi_0 - \psi_{\rm dl}} \, ,
\label{eq:sternCapacitance}
\end{equation}
where $\psi_{\rm dl}$ is the diffuse layer potential.
Finally, the charge-potential relationship closes the above set of equations:
\begin{equation}
\sigma = -\left[ 2k_{\rm B}T\varepsilon_0\varepsilon \sum_i c_i(e^{-\beta z_ie_0 \psi_{\rm dl}}-1)
\right]^{1/2} \, ,
\label{eq:chargePotentialRelationship}
\end{equation}
$\varepsilon$ is the dielectric constant, $\varepsilon_0$ is the vacuum permittivity, 
$c_i$ is the ion concentration, and $z_i$ ion valence. The following parameters were used to calculate the diffuse-layer
potential from the basic Stern model: ionization constant p$K = 7.7$, silanol groups site density
$\Gamma_0 = 4.75$~nm$^{-2}$, and Stern capacitance $C_{\rm S} = 0.12$~Fm$^{-2}$.

\section{Analysis of the Force Curves}

The force measurements are done in a sphere-sphere, while the calculations in a plate-plate geometry. The 
Derjaguin approximation is used for the transformation between these two geometries
\begin{equation}
F = 2\pi R_{\rm eff} W \, ,
\label{eq:Derjaguin}
\end{equation}
where $F$ is the force between the two spherical particles, $W$ is the energy per unit area in the plate-plate
geometry, and $R_{\rm eff}$ is the effective radius, which is equal to $R/2$ for particles with radii $R$.

The forces are modelled within DLVO theory
\begin{equation}
F = F_{\rm vdW} + F_{\rm dl} \, ,
\label{eq:DLVO}
\end{equation}
where $F_{\rm vdW}$ is the van der Waals and $F_{\rm dl}$ is the double-layer force. The former is calculated 
with non-retarded expression
\begin{equation}
F_{\rm vdW} = -\frac{HR}{12}\cdot \frac{1}{h^2} \, ,
\label{eq:vdW}
\end{equation}
where $H$ is the Hamaker constant and $h$ is the surface-surface separation.

The double-layer force is calculated by solving the Poisson-Boltzmann equation in the plate-plate geometry
\begin{equation}
\frac{{\rm d}^2 \psi(x)}{{\rm d} x^2} =  -\frac{e_0}{\varepsilon\varepsilon_0}\sum_i c_i \,\rme^{-\beta z_ie_0
\psi(x)} \, ,
\label{eq:PB}
\end{equation}
where $\psi(x)$ is the electric potential, and $x$ is the coordinate normal to the plates. The 
plates are positioned at $x = -h/2$ and $x = h/2$. The PB equation can be solved only in the $0 \le x \le h/2$ 
half-space due to symmetry. The constant regulation (CR) boundary conditions are used
\begin{gather}
\left. \frac{\mathrm{d} \psi }{\mathrm{d} x} \right|_{x=0} = 0 \quad {\rm and} \\
\left. \frac{\mathrm{d} \psi }{\mathrm{d} x} \right|_{x=h/2} = \sigma - C_{\rm in}[\psi(h/2) - \psi_{\rm dl}] 
\, ,
\end{gather}
where $\sigma$ and $\psi_{\rm dl}$ are surface charge density and diffuse layer potential of the isolated
surface, and $C_{\rm in}$ is the inner layer capacitance. Instead of using inner layer capacitance we 
introduce regulation parameter as
\begin{equation}
p = \frac{C_{\rm dl}}{C_{\rm dl} + C_{\rm in}} \, ,
\label{eq:reg_param}
\end{equation}
where diffuse layer capacitance is defined as
\begin{equation}
C_{\rm dl} =  \frac{\partial \sigma }{\partial \psi_{\rm dl}} = \left( 
\frac{e_0^2\varepsilon\varepsilon_0}{2 \kB T}\right)^{1/2}\cdot
\frac{\sum_i z_ic_i(e^{-\beta z_ie_0 \psi_{\rm dl}}-1)}
{[\sum_i c_i(e^{-\beta z_ie_0 \psi_{\rm dl}}-1)]^{1/2}}\, .
\label{eq:dl_capacitance}
\end{equation}
Regulation parameter enables to easily interpret the boundary conditions, $p = 1$ represents constant charge  
(CC) conditions, while $p = 0$ represents constant potential (CP) conditions.

The disjoining pressure is then calculated using a potential at the mid-plane $\psi(0) = \psi_{\rm M}$
\begin{equation}
\Pi = \kB T \sum_i c_i \left( e^{-z_ie_0\beta \psi_{\rm M}} - 1 \right) \, .
\end{equation}
The integration of the pressure profile results in the energy per unit area
\begin{equation}
W_{\rm dl} = \int_h^{\infty} \Pi (h') {\rm  d} h' \, .
\end{equation}
Derjaguin approximation \Eq~(\ref{eq:Derjaguin}) is used to calculate the double-layer force, 
$F_{\rm dl}$, from energy per unit area, $W_{\rm dl}$, and the total force is calculated via \Eq~(\ref{eq:DLVO}).

The PB equation is solved numerically and the solution is modelled as a mixture of 1:1 electrolyte stemming 
from pH adjustment with KOH and 1:$z$ electrolyte for the respective added salt.

For comparison and easier interpretation the Debye-H\"uckel (DH) theory is also used to calculate far-field approximation of 
the double layer force
\begin{equation}
F_{\rm dl}^{\rm DH} = 2\pi R \varepsilon\varepsilon_0 \kappa \psi_{\rm eff}^2 e^{-\kappa h}    \, ,
\end{equation}
where $\psi_{\rm eff}$ is the effective potential and 
$\kappa=\sqrt{\frac{2\beta e_0^2 I}{\varepsilon\varepsilon_0}}$ is the inverse Debye length, where $I$ is the 
ionic strength calculated as $I=\frac{1}{2}\sum_i c_iz_i^2$. For 1:$z$ electrolyte the ionic strength is 
$I =\frac{z(z+1)}{2}\cdot c_{\rm salt}$, while monovalent counterion concentration is proportional to the valence as
$zc_{\rm salt}$, where $c_{\rm salt}$ is the concentration of 1:$z$ salt.

Finally, experimental force profiles are interpreted by fitting DLVO theory and extracting the following 
parameters: salt concentration, diffuse-layer potential, regulation parameter, and Hamaker constant. In all
cases the difference between the fitted and nominal salt concentration is typically below 10-15~\%. Note that in the present
case of 1:$z$ salts, the sign of the diffuse-layer potential of silica particles can be unambiguously determined
from the force profile, since the curves would have a different shape if the surfaces would have been positively charged.
Variations of diffuse-layer potentials and regulation parameters for different pairs of particles at the same
conditions are typically between 10 and 20~\%.

\section{Results and Discussion}

Forces between pairs of silica particles were measured with colloidal probe technique based on AFM. These
forces were measured in the presence of KCl, \kk , \kkk , and \kkkk\ at pH 10.

\begin{figure}[ht]
\centering
\includegraphics[width=8.5cm]{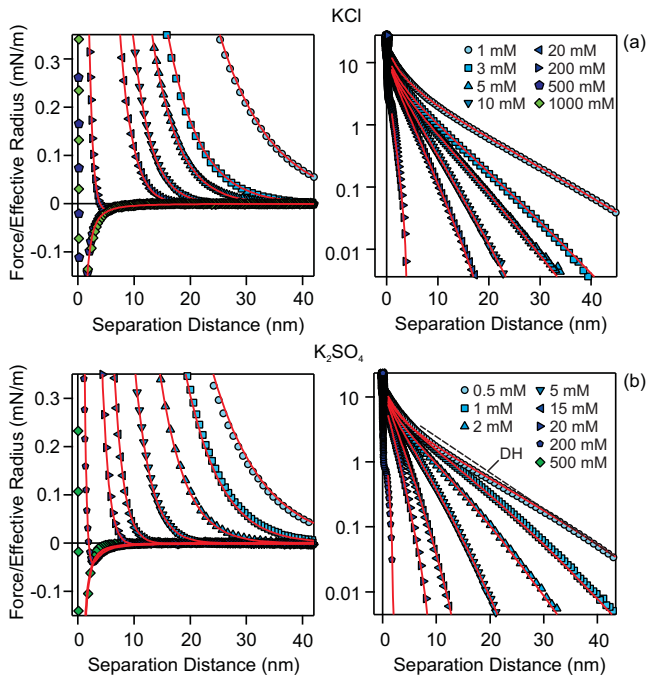}
\caption{Force curves between silica particles at different salt concentrations for (a) KCl and (b) \kk\ at pH 
10. The DLVO calculations are presented as full lines. Lin-lin representation is shown left and log-lin 
representation right. DH approximation is also shown for 0.5~mM of \kk .}
\label{fig:forcesk1k2}
\end{figure}

Forces measured between silica particles in KCl are shown in Fig.~\ref{fig:forcesk1k2}a.
On the left and right
panel forces are plotted in linear and logarithmic representations, respectively. At low salt concentrations
the forces are repulsive and they decay exponentially as expected for double-layer interactions. With
increasing salt concentration forces become shorter-ranged and finally at high concentrations above 500~mM
they turn attractive as the vdW force becomes dominant. In the log-lin representation one can also observe
that the slope of the forces in increasing with increasing concentration as the decay length shortens. DLVO
fits are shown as lines in the Fig.~\ref{fig:forcesk1k2}. The fitting strategy is the following. First the
Hamaker constant is fitted at KCl concentrations above 200~mM. An average value of 
$H = 2.6\pm 0.3\cdot 10^{-21}$~J is obtained, which is in agreement with our earlier 
results~\cite{Valmacco2016a}. The measured Hamaker constant is also close to the theoretical estimate of 
$1.6\cdot 10^{-21}$~J calculated from accurate dielectric spectra~\cite{Ackler1996}. This high value of the measured Hamaker 
constant is due to extremely small surface roughness of the silica particles heated at 1200~\celsius~\cite{Valmacco2016a}.
The fitted value of the Hamaker constant is fixed for lower concentrations and it is 
also consistent with all other used salts. For the double-layer component of the force curve, the
background 1:1 electrolyte concentration is fixed to 0.1~mM. The remaining fitting parameters are: $1:z$ salt
concentration, diffuse-layer potential, and regulation parameter. The same fitting procedure is used also for other 
salts. The fitted curves are presented as lines in
Fig.~\ref{fig:forcesk1k2}. The DLVO theory describes the force curves perfectly, except at very short 
separations below few nanometers, where the experimental curves are more repulsive then predicted by the 
theory. This short-range repulsion is probably due to the hydration forces
~\cite{Vigil1994, Valle-Delgado2005, Grabbe1993, Acuna2011} or overlapping hairy layers of polysilicilic acid~
\cite{Kobayashi2005} and it is not part of our theoretical 
description.

In Fig.~\ref{fig:forcesk1k2}b forces in the presence of \kk\ are shown. Similar behavior as in the case of
KCl is observed. The forces are repulsive at low salt concentration and become attractive at high levels of
salt. However, the onset of attractive vdW force is observed at lower concentrations as compared to the KCl 
case. Again the DLVO theory fits the data very well. The double-layer forces in the presence of divalent coions
are not exponential anymore as evident for the log-lin presentation. The exponential Debye-H\"uckel curve is
presented with the dashed line for the lowest concentration, see Fig~\ref{fig:forcesk1k2}b right. One can
observe that the experimental force at 0.5 mM of \kk\ is only exponential at distances beyond 30 nm. At
smaller separations the force deviates from the exponential behavior. Such long-range sigmoidal curves 
in the presence of mulitvalent coions were already observed by some of us~\cite{MontesRuiz-Cabello2014b} and they can become 
extremely non-exponential for the coions with large effective charge~\cite{Moazzami-Gudarzi2016c, Moazzami-Gudarzi2017}.
The source of this behavior is the exclusion
of the multivalent coions from the area between the charged surfaces at close proximity. At large distances
both monovalent counterions and mulitvalent coions are in the slit between two charged surfaces and the force
between the respective surfaces is exponential with the decay length corresponding to the inverse Debye
length for $1:z$ electrolyte. When the surfaces approach, the multivalent coions feel strong electrostatic
repulsion from the charged surfaces and get excluded from the slit. At this point only the monovalent
counterions are left in the slit. Such system behaves as salt-free (counterions-only) and results in the
power-law decay of the force as first proposed by Langmuir~\cite{Langmuir1938}. Further details on these non-exponential
force profiles can be found in~\cite{MontesRuiz-Cabello2014b}.

\begin{figure}[tbh]
\centering
\includegraphics[width=8.5cm]{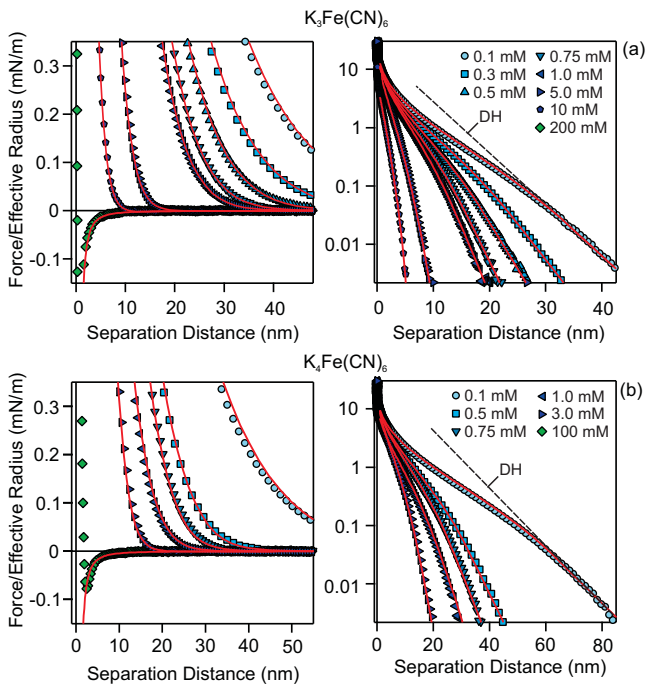}
\caption{Force curves between silica particles at different salt concentrations for (a) \kkk\ and (b) \kkkk\ 
at pH 10. The DLVO calculations are presented as full lines. Lin-lin representation is shown left and log-lin 
representation right. DH approximation is shown for the two lowest concentration.}
\label{fig:forcesk3k4}
\end{figure}

Forces measured in solutions of \kkk\ and \kkkk\ are presented in Fig.~\ref{fig:forcesk3k4}.
Subfigures~\ref{fig:forcesk3k4}a and b show the trivalent and tetravalent coion case, respectively. For both
cases forces are repulsive at low salt and attractive at high salt levels. The DLVO 
theory, shown as full lines, nicely fits the experimental data. Again, for the two lowest concentrations in the 
log-lin representations the DH approximation is presented with dashed lines. The deviation from the non-exponential 
behavior is even more evident as for the \kk\ case. Furthermore, this deviation is shifted to 
larger distances with increasing valence of the coions. The exclusion of the coions happens at larger 
distances because the repulsion between charged surfaces and coions is increasing with increasing 
valence~\cite{Trefalt2016b}.

\begin{figure*}[t]
\centering
\includegraphics[width=16cm]{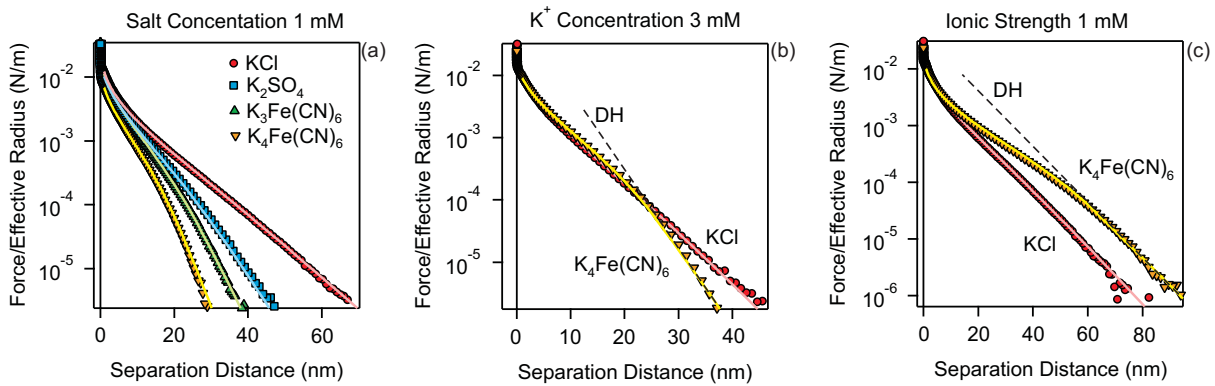}
\caption{Force curves between silica particles at (a) salt concentration of 1~mM, (b) K$^+$ concentration of 
3~mM, and (c) ionic strength of 1~mM for different salts at pH  10. In (b) and (c) only KCl and \kkkk\ are 
shown for clarity. The DLVO fits are shown as full lines, and DH approximation in (c) as dashed line.}
\label{fig:forcesComparison}
\end{figure*}

To get a further insight in the interactions between silica particles in the presence of multivalent coions we 
compare the forces for different coion valences at the same (a) salt concentration, (b) counterion 
concentration, and (c) ionic strength in Fig.~\ref{fig:forcesComparison}. 
For all the cases conditions are chosen such that the vdW force is negligible and only the double-layer component of
the force is present.

At constant salt concentration of 1~mM all forces for coion valences between 1 and 4 are repulsive, see
Fig.~\ref{fig:forcesComparison}a. Two
features can be observed. First, only the force for monovalent coion, namely Cl$^-$, is exponential, while
the interactions for the multivalent coions are non-exponential. Second, the forces get progressively
screened by increasing coion valence. This behavior is due to the increase of both the ionic strength as
well as counterion (K$^+$) concentration with increasing coion valence when the salt concentration is fixed.

If we now compare the force for
KCl and \kkkk\ at constant counterion concentration of 3 mM of K$^+$, an interesting behavior is revealed.
At large separation distances the \kkkk\ repulsion is weaker as compared to the KCl case, while at distances
below $\sim 25$~nm the two forces are comparable. As noted earlier large-distance behavior can be described
by DH approximation, therefore the force is exponential with its decay length being $\kappa^{-1}$. Since in
the present case of constant counterion concentration the ionic strength for \kkkk\ salt is higher than for
KCl, the \kkkk\ force decays faster at large separations. Also the slope in log-lin plot, which repersents 
this decay length is bigger for the tetravalent case. At smaller separations where coions are expelled the 
force is determined only by counterions in this case K$^+$ and the two force curves collapse.

The third case, presented in Fig.~\ref{fig:forcesComparison}c, shows forces at constant ionic strength of
1~mM. Here the long-distance behavior is similar for both KCl and \kkkk\ salts. Since, the ionic strength is
constant the decay length of the exponential DH behavior is the same, resulting in the same slope for both
curves in the log-lin representation. On the other side, the short-distance behavior is different, while the
KCl force stays exponential down to few nanometers, the \kkkk\ force does not decay exponentially at short
distances. Only at very short distances below $\sim 5$~nm, where tetravalent as well as monovalent coions are
expelled from the slit the two curves collapse. This situation is again dominated by counterions.

\begin{figure*}[t]
\centering
\includegraphics[width=16cm]{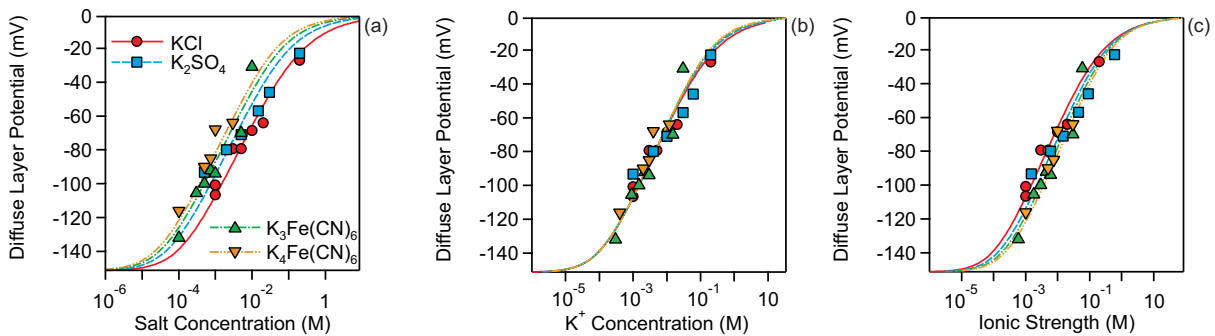}
\caption{Diffuse layer potential extracted form force curves as a function of (a) salt concentration, (b)
K$^+$ concentration, and (c) ionic strength for different salts at pH 10. The results form the basic Stern 
model are shown with lines. The parameters are: ionization constant p$K = 7.7$, silanol groups site density
$\Gamma_0 = 4.75$~nm$^{-2}$, and Stern capacitance $C_{\rm S} = 0.12$~Fm$^{-2}$.}
\label{fig:potentials}
\end{figure*}

Let us now look at the diffuse-layer potentials extracted from the force curves. In Fig.~\ref{fig:potentials}
the potentials are shown for all salts at different conditions.
The results for coions of different valences are presented at: (a) the same salt concentration, (b) the same
K$^+$ concentration, and (c) the same ionic strength. The results of basic Stern model are shown as lines
for comparison. In all cases the diffuse-layer potential is negative at low
concentrations due to charged silanol groups on the silica surface, see Eq.~\ref{eq:silanol}. With increasing 
concentration the potential increases and is neutralized at very high salt levels. The diffuse-layer potential curves
are shifted to lower concentrations when the coion valence is increased, see Fig.~\ref{fig:potentials}a.
Although there is some scatter
in the experimental data they nicely follow the basic Stern model. The presented coion valence trend can be 
rationalized in the following way. At constant salt concentration, the K$^+$ concentration increases as 
$zc_{\rm salt}$, the surface charge is screened more strongly when K$^+$ concentration is increased, and
this leads to lower magnitude of the diffuse-layer potential. This rationale is confirmed by Fig.~\ref{fig:potentials}b
where the potentials are plotted as a function of the K$^+$ concentration. All the experimental data as
well as basic Stern model calculations collapse on a single curve, showing that the potential is only a
function of the counterion concentration. In a recent study by Trompette~\cite{Trompette2017}, the type of monovalent
coion is shown to have an effect on silica nanoparicle aggregation. This behavior suggests specific adsorption of coions
to the silica surface, on the contrary our results suggest no specific adsorption of the coions. In our case the coions are
multivalent and feel stronger repulsion from the silica surface which probably hinders their adsorption.
In Fig.~\ref{fig:potentials}c the diffuse-layer potential is plotted as a
function of ionic strength. Here the trend is reversed, the potential curves are shifted to higher ionic 
strength as coion valence increases, however in this case the trend is weaker. Similar reversal is observed
in stability of colloidal suspensions as a function of coion valence and is referred to as the inverse Schulze-Hardy
rule~\cite{Cao2015, Trefalt2017a}. 

\begin{figure*}[t]
\centering
\includegraphics[width=16cm]{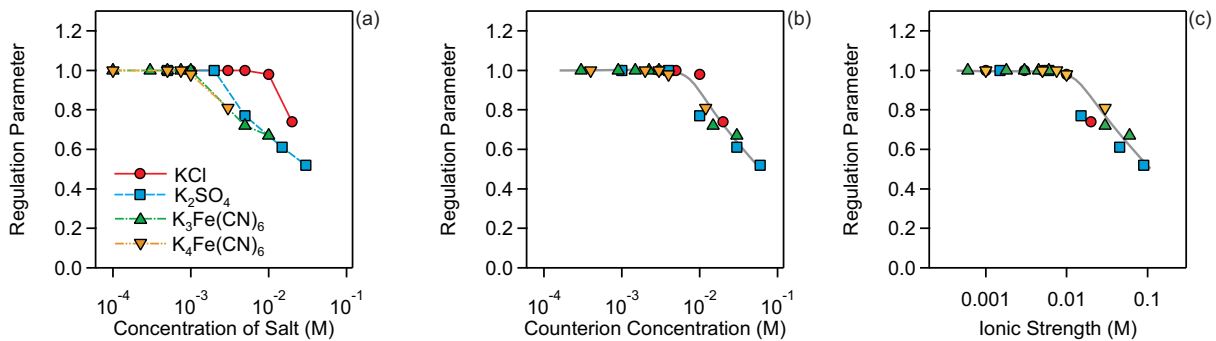}
\caption{Regulation parameter extracted form force curves as a function of (a) salt concentration, (b)
K$^+$ concentration, and (c) ionic strength for different salts at pH 10. Lines are only used to guide the eye.}
\label{fig:regulation}
\end{figure*}

Finally we examine another important property of charged surfaces, namely its regulation behavior.
In Fig.~\ref{fig:regulation} regulation parameter is presented in a similar manner as the diffuse-layer potential above.
Regulation parameter describes how the charge on the surface is changing upon approach of the two particles.
If regulation parameter, $p$, is close to unity, the charge on the surface is constant upon approach. On the
other hand for $p < 1$, the surface regulates/adjusts its charge upon approach~\cite{Trefalt2016}.
Fig.~\ref{fig:regulation} reveals that the general trend for all salts
is the same, regulation parameter is close to unity at low salt concentration and it decreases with increasing
concentration. One would expect an increase in regulation parameter in a situation where the inner layer
capacitance, $C_{\rm in}$, is constant or it is decreasing with increasing concentration~\cite{Trefalt2016}.
However, it has been recently shown with X-ray photoelectron spectroscopy in a liquid microjet, that the
inner layer capacitance of silica surface is increasing with increasing concentration~\cite{Brown2016}. If $C_{\rm in}$
is increasing with concentration the regulation parameter decreases with concentration, see Eq.~(\ref{eq:reg_param}).
Such behavior is in line with our observations of decreasing regulation parameter with increasing concentration.
At low concentrations below about 1~mM we have fixed $p = 1$ in our fits in order to avoid the values above unity 
which are not consistent with the regulation model. Due to this modification the measured forces were slightly more repulsive
than the calculated ones at short distances, and this behavior if probably connected to the short-range hydration repulsion 
which is known to be present for silica surfaces in aqueous solutions~\cite{Vigil1994, Valle-Delgado2005, Grabbe1993, Acuna2011, Valmacco2016}.

Similarly to the diffuse-layer potential, regulation parameter curves also shift to lower salt concentrations 
with increasing coion valence, see Fig.~\ref{fig:regulation}a. When the $p$ is plotted as a function of
either K$^+$ concentration or ionic strength the data points collapse on a single curve. Since there is  
relatively large scatter in the experimental points it is not possible to determine which collapse is
correct. Regulation behavior is determined by adsorption of K$^+$ ions to the surface
and therefore the collapse on K$^+$ concentration is probably a relevant one.

\section{Conclusions}

We have measured forces between negatively charged silica particles in the presence of 1:$z$ salts. In these systems the
multivalent ions have the same charge as the surface and therefore play the role of coions. The measured forces are
repulsive at low concentrations where the double-layer forces are strong, while at high concentration the attractive vdW
forces are dominant as the electrostatic interactions are screened away. The double-layer force profiles assume a
non-exponential shape, where transition between a long-range exponential and short-range power-law behavior is observed.
This transition is due to the electrostatic exclusion of multivalent coions from the slit at smaller separation distances.
At large separation distances the forces are parallel at constant ionic strength if plotted in log-lin representation.
On the other hand at closer separations the forces overlap for different coion valences 
if the counterion concentration is constant.
This behavior suggests that the monovalent counterion concentration determines the near-field force. The diffuse-layer
potentials are increasing with increasing concentration for all investigated salts. The potential curves shift to lower
salt concentrations with increasing coion valence, since the salts with higher valence is more effective in screening the
surface charge. This behavior is in line with 1-p$K$ basic Stern model predictions. The order of the
potential curves gets reversed if they are plotted as a function of ionic strength, $i.e.$ the potential curve for lower 
valence comes first. Finally, if one plots the potentials as a function of monovalent counterion concentration all the
experimental data as well as the basic Stern model calculations collapse on a single master curve. Similar shifts to lower
salt concentrations with increasing coion valence are also observed for the regulation parameter. Furthermore, the 
collapse on a single master curve is also observed if regulation parameter is plotted as a function of monovalent counterion
concentrations. From this behavior one can conclude that for a asymmetric 1:$z$ electrolytes, where $z$ represents
the coion, the double-layer properties are mainly determined by the concentration of the monovalent counterions.

\section*{Acknowledgements}
This research was supported by the Swiss National Science Foundation through grant 162420 and the University of Geneva. The 
authors are thankful to Michal Borkovec for providing access to the instruments in his laboratory and to Plinio Maroni for
the help with AFM measurements. 

\bibliography{paperLib}

\begin{thebibliography}{47}%
\makeatletter
\providecommand \@ifxundefined [1]{%
 \@ifx{#1\undefined}
}%
\providecommand \@ifnum [1]{%
 \ifnum #1\expandafter \@firstoftwo
 \else \expandafter \@secondoftwo
 \fi
}%
\providecommand \@ifx [1]{%
 \ifx #1\expandafter \@firstoftwo
 \else \expandafter \@secondoftwo
 \fi
}%
\providecommand \natexlab [1]{#1}%
\providecommand \enquote  [1]{``#1''}%
\providecommand \bibnamefont  [1]{#1}%
\providecommand \bibfnamefont [1]{#1}%
\providecommand \citenamefont [1]{#1}%
\providecommand \href@noop [0]{\@secondoftwo}%
\providecommand \href [0]{\begingroup \@sanitize@url \@href}%
\providecommand \@href[1]{\@@startlink{#1}\@@href}%
\providecommand \@@href[1]{\endgroup#1\@@endlink}%
\providecommand \@sanitize@url [0]{\catcode `\\12\catcode `\$12\catcode
  `\&12\catcode `\#12\catcode `\^12\catcode `\_12\catcode `\%12\relax}%
\providecommand \@@startlink[1]{}%
\providecommand \@@endlink[0]{}%
\providecommand \url  [0]{\begingroup\@sanitize@url \@url }%
\providecommand \@url [1]{\endgroup\@href {#1}{\urlprefix }}%
\providecommand \urlprefix  [0]{URL }%
\providecommand \Eprint [0]{\href }%
\providecommand \doibase [0]{http://dx.doi.org/}%
\providecommand \selectlanguage [0]{\@gobble}%
\providecommand \bibinfo  [0]{\@secondoftwo}%
\providecommand \bibfield  [0]{\@secondoftwo}%
\providecommand \translation [1]{[#1]}%
\providecommand \BibitemOpen [0]{}%
\providecommand \bibitemStop [0]{}%
\providecommand \bibitemNoStop [0]{.\EOS\space}%
\providecommand \EOS [0]{\spacefactor3000\relax}%
\providecommand \BibitemShut  [1]{\csname bibitem#1\endcsname}%
\let\auto@bib@innerbib\@empty
\bibitem [{\citenamefont {Strey}\ \emph {et~al.}(1998)\citenamefont {Strey},
  \citenamefont {Podgornik}, \citenamefont {Rau},\ and\ \citenamefont
  {Parsegian}}]{Strey1998}%
  \BibitemOpen
  \bibfield  {author} {\bibinfo {author} {\bibfnamefont {H.~H.}\ \bibnamefont
  {Strey}}, \bibinfo {author} {\bibfnamefont {R.}~\bibnamefont {Podgornik}},
  \bibinfo {author} {\bibfnamefont {D.~C.}\ \bibnamefont {Rau}}, \ and\
  \bibinfo {author} {\bibfnamefont {V.~A.}\ \bibnamefont {Parsegian}},\ }\href
  {\doibase 10.1016/S0959-440X(98)80063-8} {\bibfield  {journal} {\bibinfo
  {journal} {Curr. Opin. Struct. Biol.}\ }\textbf {\bibinfo {volume} {8}},\
  \bibinfo {pages} {309} (\bibinfo {year} {1998})}\BibitemShut {NoStop}%
\bibitem [{\citenamefont {Iselau}\ \emph {et~al.}(2015)\citenamefont {Iselau},
  \citenamefont {Restorp}, \citenamefont {Andersson},\ and\ \citenamefont
  {Bordes}}]{Iselau2015}%
  \BibitemOpen
  \bibfield  {author} {\bibinfo {author} {\bibfnamefont {F.}~\bibnamefont
  {Iselau}}, \bibinfo {author} {\bibfnamefont {P.}~\bibnamefont {Restorp}},
  \bibinfo {author} {\bibfnamefont {M.}~\bibnamefont {Andersson}}, \ and\
  \bibinfo {author} {\bibfnamefont {R.}~\bibnamefont {Bordes}},\ }\href
  {\doibase 10.1016/j.colsurfa.2015.04.013} {\bibfield  {journal} {\bibinfo
  {journal} {Colloids Surf., A}\ }\textbf {\bibinfo {volume} {483}},\ \bibinfo
  {pages} {264} (\bibinfo {year} {2015})}\BibitemShut {NoStop}%
\bibitem [{\citenamefont {Bolto}\ and\ \citenamefont
  {Gregory}(2007)}]{Bolto2007}%
  \BibitemOpen
  \bibfield  {author} {\bibinfo {author} {\bibfnamefont {B.}~\bibnamefont
  {Bolto}}\ and\ \bibinfo {author} {\bibfnamefont {J.}~\bibnamefont
  {Gregory}},\ }\href@noop {} {\bibfield  {journal} {\bibinfo  {journal} {Water
  Research}\ }\textbf {\bibinfo {volume} {41}},\ \bibinfo {pages} {2301}
  (\bibinfo {year} {2007})}\BibitemShut {NoStop}%
\bibitem [{\citenamefont {Cerbelaud}\ \emph {et~al.}(2010)\citenamefont
  {Cerbelaud}, \citenamefont {Videcoq}, \citenamefont {Abelard}, \citenamefont
  {Pagnoux}, \citenamefont {Rossignol},\ and\ \citenamefont
  {Ferrando}}]{Cerbelaud2010}%
  \BibitemOpen
  \bibfield  {author} {\bibinfo {author} {\bibfnamefont {M.}~\bibnamefont
  {Cerbelaud}}, \bibinfo {author} {\bibfnamefont {A.}~\bibnamefont {Videcoq}},
  \bibinfo {author} {\bibfnamefont {P.}~\bibnamefont {Abelard}}, \bibinfo
  {author} {\bibfnamefont {C.}~\bibnamefont {Pagnoux}}, \bibinfo {author}
  {\bibfnamefont {F.}~\bibnamefont {Rossignol}}, \ and\ \bibinfo {author}
  {\bibfnamefont {R.}~\bibnamefont {Ferrando}},\ }\href@noop {} {\bibfield
  {journal} {\bibinfo  {journal} {Soft Matter}\ }\textbf {\bibinfo {volume}
  {6}},\ \bibinfo {pages} {370} (\bibinfo {year} {2010})}\BibitemShut {NoStop}%
\bibitem [{\citenamefont {Kuscer}\ \emph {et~al.}(2012)\citenamefont {Kuscer},
  \citenamefont {Stavber}, \citenamefont {Trefalt},\ and\ \citenamefont
  {Kosec}}]{Kuscer2012}%
  \BibitemOpen
  \bibfield  {author} {\bibinfo {author} {\bibfnamefont {D.}~\bibnamefont
  {Kuscer}}, \bibinfo {author} {\bibfnamefont {G.}~\bibnamefont {Stavber}},
  \bibinfo {author} {\bibfnamefont {G.}~\bibnamefont {Trefalt}}, \ and\
  \bibinfo {author} {\bibfnamefont {M.}~\bibnamefont {Kosec}},\ }\href
  {\doibase 10.1111/j.1551-2916.2011.04876.x} {\bibfield  {journal} {\bibinfo
  {journal} {J. Am. Ceram. Soc.}\ }\textbf {\bibinfo {volume} {95}},\ \bibinfo
  {pages} {487} (\bibinfo {year} {2012})}\BibitemShut {NoStop}%
\bibitem [{\citenamefont {Zanini}\ \emph {et~al.}(2017)\citenamefont {Zanini},
  \citenamefont {Hsu}, \citenamefont {Magrini}, \citenamefont {Marini},\ and\
  \citenamefont {Isa}}]{Zanini2017}%
  \BibitemOpen
  \bibfield  {author} {\bibinfo {author} {\bibfnamefont {M.}~\bibnamefont
  {Zanini}}, \bibinfo {author} {\bibfnamefont {C.-P.}\ \bibnamefont {Hsu}},
  \bibinfo {author} {\bibfnamefont {T.}~\bibnamefont {Magrini}}, \bibinfo
  {author} {\bibfnamefont {E.}~\bibnamefont {Marini}}, \ and\ \bibinfo {author}
  {\bibfnamefont {L.}~\bibnamefont {Isa}},\ }\href {\doibase
  10.1016/j.colsurfa.2017.05.084} {\bibfield  {journal} {\bibinfo  {journal}
  {Colloids Surf., A}\ }\textbf {\bibinfo {volume} {in press}} (\bibinfo {year}
  {2017}),\ 10.1016/j.colsurfa.2017.05.084}\BibitemShut {NoStop}%
\bibitem [{\citenamefont {Labbez}\ \emph {et~al.}(2006)\citenamefont {Labbez},
  \citenamefont {Jonsson}, \citenamefont {Pochard}, \citenamefont {Nonat},\
  and\ \citenamefont {Cabane}}]{Labbez2006}%
  \BibitemOpen
  \bibfield  {author} {\bibinfo {author} {\bibfnamefont {C.}~\bibnamefont
  {Labbez}}, \bibinfo {author} {\bibfnamefont {B.}~\bibnamefont {Jonsson}},
  \bibinfo {author} {\bibfnamefont {I.}~\bibnamefont {Pochard}}, \bibinfo
  {author} {\bibfnamefont {A.}~\bibnamefont {Nonat}}, \ and\ \bibinfo {author}
  {\bibfnamefont {B.}~\bibnamefont {Cabane}},\ }\href@noop {} {\bibfield
  {journal} {\bibinfo  {journal} {J. Phys. Chem. B}\ }\textbf {\bibinfo
  {volume} {110}},\ \bibinfo {pages} {9219} (\bibinfo {year}
  {2006})}\BibitemShut {NoStop}%
\bibitem [{\citenamefont {Israelachvili}\ \emph {et~al.}(2010)\citenamefont
  {Israelachvili}, \citenamefont {Min}, \citenamefont {Akbulut}, \citenamefont
  {Alig}, \citenamefont {Carver}, \citenamefont {Greene}, \citenamefont
  {Kristiansen}, \citenamefont {Meyer}, \citenamefont {Pesika}, \citenamefont
  {{K Rosenberg}},\ and\ \citenamefont {Zeng}}]{Israelachvili2010}%
  \BibitemOpen
  \bibfield  {author} {\bibinfo {author} {\bibfnamefont {J.}~\bibnamefont
  {Israelachvili}}, \bibinfo {author} {\bibfnamefont {Y.}~\bibnamefont {Min}},
  \bibinfo {author} {\bibfnamefont {M.}~\bibnamefont {Akbulut}}, \bibinfo
  {author} {\bibfnamefont {A.}~\bibnamefont {Alig}}, \bibinfo {author}
  {\bibfnamefont {G.}~\bibnamefont {Carver}}, \bibinfo {author} {\bibfnamefont
  {W.}~\bibnamefont {Greene}}, \bibinfo {author} {\bibfnamefont
  {K.}~\bibnamefont {Kristiansen}}, \bibinfo {author} {\bibfnamefont
  {E.}~\bibnamefont {Meyer}}, \bibinfo {author} {\bibfnamefont
  {N.}~\bibnamefont {Pesika}}, \bibinfo {author} {\bibnamefont {{K
  Rosenberg}}}, \ and\ \bibinfo {author} {\bibfnamefont {H.}~\bibnamefont
  {Zeng}},\ }\href {\doibase 10.1088/0034-4885/73/3/036601} {\bibfield
  {journal} {\bibinfo  {journal} {Rep. Prog. Phys.}\ }\textbf {\bibinfo
  {volume} {73}},\ \bibinfo {pages} {036601} (\bibinfo {year}
  {2010})}\BibitemShut {NoStop}%
\bibitem [{\citenamefont {Espinosa-Marzal}\ \emph {et~al.}(2012)\citenamefont
  {Espinosa-Marzal}, \citenamefont {Drobek}, \citenamefont {Balmer},\ and\
  \citenamefont {Heuberger}}]{Espinosa-Marzal2012}%
  \BibitemOpen
  \bibfield  {author} {\bibinfo {author} {\bibfnamefont {R.~M.}\ \bibnamefont
  {Espinosa-Marzal}}, \bibinfo {author} {\bibfnamefont {T.}~\bibnamefont
  {Drobek}}, \bibinfo {author} {\bibfnamefont {T.}~\bibnamefont {Balmer}}, \
  and\ \bibinfo {author} {\bibfnamefont {M.~P.}\ \bibnamefont {Heuberger}},\
  }\href@noop {} {\bibfield  {journal} {\bibinfo  {journal} {Phys. Chem. Chem.
  Phys.}\ }\textbf {\bibinfo {volume} {14}},\ \bibinfo {pages} {6085} (\bibinfo
  {year} {2012})}\BibitemShut {NoStop}%
\bibitem [{\citenamefont {Ducker}\ \emph {et~al.}(1991)\citenamefont {Ducker},
  \citenamefont {Senden},\ and\ \citenamefont {Pashley}}]{Ducker1991}%
  \BibitemOpen
  \bibfield  {author} {\bibinfo {author} {\bibfnamefont {W.~A.}\ \bibnamefont
  {Ducker}}, \bibinfo {author} {\bibfnamefont {T.~J.}\ \bibnamefont {Senden}},
  \ and\ \bibinfo {author} {\bibfnamefont {R.~M.}\ \bibnamefont {Pashley}},\
  }\href@noop {} {\bibfield  {journal} {\bibinfo  {journal} {Nature}\ }\textbf
  {\bibinfo {volume} {353}},\ \bibinfo {pages} {239} (\bibinfo {year}
  {1991})}\BibitemShut {NoStop}%
\bibitem [{\citenamefont {Butt}(1991)}]{Butt1991}%
  \BibitemOpen
  \bibfield  {author} {\bibinfo {author} {\bibfnamefont {H.~J.}\ \bibnamefont
  {Butt}},\ }\href@noop {} {\bibfield  {journal} {\bibinfo  {journal} {Biophys.
  J.}\ }\textbf {\bibinfo {volume} {60}},\ \bibinfo {pages} {1438} (\bibinfo
  {year} {1991})}\BibitemShut {NoStop}%
\bibitem [{\citenamefont {Prieve}(1999)}]{Prieve1999}%
  \BibitemOpen
  \bibfield  {author} {\bibinfo {author} {\bibfnamefont {D.~C.}\ \bibnamefont
  {Prieve}},\ }\href@noop {} {\bibfield  {journal} {\bibinfo  {journal} {Adv.
  Colloid Interface Sci.}\ }\textbf {\bibinfo {volume} {82}},\ \bibinfo {pages}
  {93} (\bibinfo {year} {1999})}\BibitemShut {NoStop}%
\bibitem [{\citenamefont {{von Grunberg}}\ \emph {et~al.}(2001)\citenamefont
  {{von Grunberg}}, \citenamefont {Helden}, \citenamefont {Leiderer},\ and\
  \citenamefont {Bechinger}}]{vonGrunberg2001}%
  \BibitemOpen
  \bibfield  {author} {\bibinfo {author} {\bibfnamefont {H.~H.}\ \bibnamefont
  {{von Grunberg}}}, \bibinfo {author} {\bibfnamefont {L.}~\bibnamefont
  {Helden}}, \bibinfo {author} {\bibfnamefont {P.}~\bibnamefont {Leiderer}}, \
  and\ \bibinfo {author} {\bibfnamefont {C.}~\bibnamefont {Bechinger}},\
  }\href@noop {} {\bibfield  {journal} {\bibinfo  {journal} {J. Chem. Phys.}\
  }\textbf {\bibinfo {volume} {114}},\ \bibinfo {pages} {10094} (\bibinfo
  {year} {2001})}\BibitemShut {NoStop}%
\bibitem [{\citenamefont {Gutsche}\ \emph {et~al.}(2007)\citenamefont
  {Gutsche}, \citenamefont {Keyser}, \citenamefont {Kegler},\ and\
  \citenamefont {Kremer}}]{Gutsche2007}%
  \BibitemOpen
  \bibfield  {author} {\bibinfo {author} {\bibfnamefont {C.}~\bibnamefont
  {Gutsche}}, \bibinfo {author} {\bibfnamefont {U.~F.}\ \bibnamefont {Keyser}},
  \bibinfo {author} {\bibfnamefont {K.}~\bibnamefont {Kegler}}, \ and\ \bibinfo
  {author} {\bibfnamefont {F.}~\bibnamefont {Kremer}},\ }\href@noop {}
  {\bibfield  {journal} {\bibinfo  {journal} {Phys. Rev. E}\ }\textbf {\bibinfo
  {volume} {76}},\ \bibinfo {pages} {031403} (\bibinfo {year}
  {2007})}\BibitemShut {NoStop}%
\bibitem [{\citenamefont {Crocker}\ and\ \citenamefont
  {Grier}(1994)}]{Crocker1994}%
  \BibitemOpen
  \bibfield  {author} {\bibinfo {author} {\bibfnamefont {J.~C.}\ \bibnamefont
  {Crocker}}\ and\ \bibinfo {author} {\bibfnamefont {D.~G.}\ \bibnamefont
  {Grier}},\ }\href@noop {} {\bibfield  {journal} {\bibinfo  {journal} {Phys.
  Rev. Lett.}\ }\textbf {\bibinfo {volume} {73}},\ \bibinfo {pages} {352}
  (\bibinfo {year} {1994})}\BibitemShut {NoStop}%
\bibitem [{\citenamefont {Pashley}(1984)}]{Pashley1984}%
  \BibitemOpen
  \bibfield  {author} {\bibinfo {author} {\bibfnamefont {R.~M.}\ \bibnamefont
  {Pashley}},\ }\href@noop {} {\bibfield  {journal} {\bibinfo  {journal} {J.
  Colloid Interface Sci.}\ }\textbf {\bibinfo {volume} {102}},\ \bibinfo
  {pages} {23} (\bibinfo {year} {1984})}\BibitemShut {NoStop}%
\bibitem [{\citenamefont {{Besteman}}\ \emph {et~al.}(2004)\citenamefont
  {{Besteman}}, \citenamefont {{Zevenbergen}}, \citenamefont {{Heering}},\ and\
  \citenamefont {{Lemay}}}]{Besteman2004}%
  \BibitemOpen
  \bibfield  {author} {\bibinfo {author} {\bibfnamefont {K.}~\bibnamefont
  {{Besteman}}}, \bibinfo {author} {\bibfnamefont {M.~A.~G.}\ \bibnamefont
  {{Zevenbergen}}}, \bibinfo {author} {\bibfnamefont {H.~A.}\ \bibnamefont
  {{Heering}}}, \ and\ \bibinfo {author} {\bibfnamefont {S.~G.}\ \bibnamefont
  {{Lemay}}},\ }\href@noop {} {\bibfield  {journal} {\bibinfo  {journal} {Phys.
  Rev. Lett.}\ }\textbf {\bibinfo {volume} {93}},\ \bibinfo {pages} {170802}
  (\bibinfo {year} {2004})}\BibitemShut {NoStop}%
\bibitem [{\citenamefont {{Zohar}}\ \emph {et~al.}(2006)\citenamefont
  {{Zohar}}, \citenamefont {{Leizerson}},\ and\ \citenamefont
  {{Sivan}}}]{Zohar2006}%
  \BibitemOpen
  \bibfield  {author} {\bibinfo {author} {\bibfnamefont {O.}~\bibnamefont
  {{Zohar}}}, \bibinfo {author} {\bibfnamefont {I.}~\bibnamefont
  {{Leizerson}}}, \ and\ \bibinfo {author} {\bibfnamefont {U.}~\bibnamefont
  {{Sivan}}},\ }\href@noop {} {\bibfield  {journal} {\bibinfo  {journal} {Phys.
  Rev. Lett.}\ }\textbf {\bibinfo {volume} {96}},\ \bibinfo {pages} {177802}
  (\bibinfo {year} {2006})}\BibitemShut {NoStop}%
\bibitem [{\citenamefont {{Sinha}}\ \emph {et~al.}(2013)\citenamefont
  {{Sinha}}, \citenamefont {{Szilagyi}}, \citenamefont {{Montes Ruiz-Cabello}},
  \citenamefont {{Maroni}},\ and\ \citenamefont {{Borkovec}}}]{Sinha2013}%
  \BibitemOpen
  \bibfield  {author} {\bibinfo {author} {\bibfnamefont {P.}~\bibnamefont
  {{Sinha}}}, \bibinfo {author} {\bibfnamefont {I.}~\bibnamefont {{Szilagyi}}},
  \bibinfo {author} {\bibfnamefont {F.~J.}\ \bibnamefont {{Montes
  Ruiz-Cabello}}}, \bibinfo {author} {\bibfnamefont {P.}~\bibnamefont
  {{Maroni}}}, \ and\ \bibinfo {author} {\bibfnamefont {M.}~\bibnamefont
  {{Borkovec}}},\ }\href@noop {} {\bibfield  {journal} {\bibinfo  {journal} {J.
  Phys. Chem. Lett.}\ }\textbf {\bibinfo {volume} {4}},\ \bibinfo {pages} {648}
  (\bibinfo {year} {2013})}\BibitemShut {NoStop}%
\bibitem [{\citenamefont {Danov}\ \emph {et~al.}(2016)\citenamefont {Danov},
  \citenamefont {Basheva},\ and\ \citenamefont {Kralchevsky}}]{Danov2016}%
  \BibitemOpen
  \bibfield  {author} {\bibinfo {author} {\bibfnamefont {K.~D.}\ \bibnamefont
  {Danov}}, \bibinfo {author} {\bibfnamefont {E.~S.}\ \bibnamefont {Basheva}},
  \ and\ \bibinfo {author} {\bibfnamefont {P.~A.}\ \bibnamefont
  {Kralchevsky}},\ }\href {\doibase 10.3390/ma9030145} {\bibfield  {journal}
  {\bibinfo  {journal} {Materials}\ }\textbf {\bibinfo {volume} {9}},\ \bibinfo
  {pages} {145} (\bibinfo {year} {2016})}\BibitemShut {NoStop}%
\bibitem [{\citenamefont {{Montes Ruiz-Cabello}}\ \emph
  {et~al.}(2014)\citenamefont {{Montes Ruiz-Cabello}}, \citenamefont
  {{Trefalt}}, \citenamefont {{Maroni}},\ and\ \citenamefont
  {{Borkovec}}}]{MontesRuiz-Cabello2014b}%
  \BibitemOpen
  \bibfield  {author} {\bibinfo {author} {\bibfnamefont {F.~J.}\ \bibnamefont
  {{Montes Ruiz-Cabello}}}, \bibinfo {author} {\bibfnamefont {G.}~\bibnamefont
  {{Trefalt}}}, \bibinfo {author} {\bibfnamefont {P.}~\bibnamefont {{Maroni}}},
  \ and\ \bibinfo {author} {\bibfnamefont {M.}~\bibnamefont {{Borkovec}}},\
  }\href {\doibase 10.1021/la500612a} {\bibfield  {journal} {\bibinfo
  {journal} {{Langmuir}}\ }\textbf {\bibinfo {volume} {30}},\ \bibinfo {pages}
  {4551} (\bibinfo {year} {2014})}\BibitemShut {NoStop}%
\bibitem [{\citenamefont {{Moazzami Gudarzi}}\ \emph
  {et~al.}(2015)\citenamefont {{Moazzami Gudarzi}}, \citenamefont {{Trefalt}},
  \citenamefont {{Szilagyi}}, \citenamefont {{Maroni}},\ and\ \citenamefont
  {{Borkovec}}}]{MoazzamiGudarzi2015}%
  \BibitemOpen
  \bibfield  {author} {\bibinfo {author} {\bibfnamefont {M.}~\bibnamefont
  {{Moazzami Gudarzi}}}, \bibinfo {author} {\bibfnamefont {G.}~\bibnamefont
  {{Trefalt}}}, \bibinfo {author} {\bibfnamefont {I.}~\bibnamefont
  {{Szilagyi}}}, \bibinfo {author} {\bibfnamefont {P.}~\bibnamefont
  {{Maroni}}}, \ and\ \bibinfo {author} {\bibfnamefont {M.}~\bibnamefont
  {{Borkovec}}},\ }\href {\doibase 10.1021/acs.jpcc.5b04426} {\bibfield
  {journal} {\bibinfo  {journal} {J. Phys. Chem. C}\ }\textbf {\bibinfo
  {volume} {119}},\ \bibinfo {pages} {15482} (\bibinfo {year}
  {2015})}\BibitemShut {NoStop}%
\bibitem [{\citenamefont {Valmacco}\ \emph
  {et~al.}(2016{\natexlab{a}})\citenamefont {Valmacco}, \citenamefont
  {Elzbieciak-Wodka}, \citenamefont {Herman}, \citenamefont {Trefalt},
  \citenamefont {Maroni},\ and\ \citenamefont {Borkovec}}]{Valmacco2016}%
  \BibitemOpen
  \bibfield  {author} {\bibinfo {author} {\bibfnamefont {V.}~\bibnamefont
  {Valmacco}}, \bibinfo {author} {\bibfnamefont {M.}~\bibnamefont
  {Elzbieciak-Wodka}}, \bibinfo {author} {\bibfnamefont {D.}~\bibnamefont
  {Herman}}, \bibinfo {author} {\bibfnamefont {G.}~\bibnamefont {Trefalt}},
  \bibinfo {author} {\bibfnamefont {P.}~\bibnamefont {Maroni}}, \ and\ \bibinfo
  {author} {\bibfnamefont {M.}~\bibnamefont {Borkovec}},\ }\href {\doibase
  10.1016/j.jcis.2016.03.043} {\bibfield  {journal} {\bibinfo  {journal} {J.
  Colloid Interface Sci.}\ }\textbf {\bibinfo {volume} {472}},\ \bibinfo
  {pages} {108} (\bibinfo {year} {2016}{\natexlab{a}})}\BibitemShut {NoStop}%
\bibitem [{\citenamefont {Trefalt}\ \emph
  {et~al.}(2017{\natexlab{a}})\citenamefont {Trefalt}, \citenamefont
  {Palberg},\ and\ \citenamefont {Borkovec}}]{Trefalt2017a}%
  \BibitemOpen
  \bibfield  {author} {\bibinfo {author} {\bibfnamefont {G.}~\bibnamefont
  {Trefalt}}, \bibinfo {author} {\bibfnamefont {T.}~\bibnamefont {Palberg}}, \
  and\ \bibinfo {author} {\bibfnamefont {M.}~\bibnamefont {Borkovec}},\ }\href
  {\doibase 10.1016/j.cocis.2016.09.008} {\bibfield  {journal} {\bibinfo
  {journal} {Curr. Opin. Colloid Interface Sci.}\ }\textbf {\bibinfo {volume}
  {27}},\ \bibinfo {pages} {9} (\bibinfo {year}
  {2017}{\natexlab{a}})}\BibitemShut {NoStop}%
\bibitem [{\citenamefont {{Trulsson}}\ \emph {et~al.}(2006)\citenamefont
  {{Trulsson}}, \citenamefont {{Jonsson}}, \citenamefont {{Akesson}},
  \citenamefont {{Forsman}},\ and\ \citenamefont {{Labbez}}}]{Trulsson2006}%
  \BibitemOpen
  \bibfield  {author} {\bibinfo {author} {\bibfnamefont {M.}~\bibnamefont
  {{Trulsson}}}, \bibinfo {author} {\bibfnamefont {B.}~\bibnamefont
  {{Jonsson}}}, \bibinfo {author} {\bibfnamefont {T.}~\bibnamefont
  {{Akesson}}}, \bibinfo {author} {\bibfnamefont {J.}~\bibnamefont
  {{Forsman}}}, \ and\ \bibinfo {author} {\bibfnamefont {C.}~\bibnamefont
  {{Labbez}}},\ }\href@noop {} {\bibfield  {journal} {\bibinfo  {journal}
  {Phys. Rev. Lett.}\ }\textbf {\bibinfo {volume} {97}},\ \bibinfo {pages}
  {068302} (\bibinfo {year} {2006})}\BibitemShut {NoStop}%
\bibitem [{\citenamefont {Kandu{\v c}}\ \emph {et~al.}(2017)\citenamefont
  {Kandu{\v c}}, \citenamefont {Moazzami-Gudarzi}, \citenamefont {Valmacco},
  \citenamefont {Podgornik},\ and\ \citenamefont {Trefalt}}]{Kanduc2017}%
  \BibitemOpen
  \bibfield  {author} {\bibinfo {author} {\bibfnamefont {M.}~\bibnamefont
  {Kandu{\v c}}}, \bibinfo {author} {\bibfnamefont {M.}~\bibnamefont
  {Moazzami-Gudarzi}}, \bibinfo {author} {\bibfnamefont {V.}~\bibnamefont
  {Valmacco}}, \bibinfo {author} {\bibfnamefont {R.}~\bibnamefont {Podgornik}},
  \ and\ \bibinfo {author} {\bibfnamefont {G.}~\bibnamefont {Trefalt}},\ }\href
  {\doibase 10.1039/C7CP00685C} {\bibfield  {journal} {\bibinfo  {journal}
  {Phys. Chem. Chem. Phys.}\ }\textbf {\bibinfo {volume} {19}},\ \bibinfo
  {pages} {10069} (\bibinfo {year} {2017})}\BibitemShut {NoStop}%
\bibitem [{\citenamefont {{Kohonen}}\ \emph {et~al.}(2000)\citenamefont
  {{Kohonen}}, \citenamefont {{Karaman}},\ and\ \citenamefont
  {{Pashley}}}]{Kohonen2000}%
  \BibitemOpen
  \bibfield  {author} {\bibinfo {author} {\bibfnamefont {M.~M.}\ \bibnamefont
  {{Kohonen}}}, \bibinfo {author} {\bibfnamefont {M.~E.}\ \bibnamefont
  {{Karaman}}}, \ and\ \bibinfo {author} {\bibfnamefont {R.~M.}\ \bibnamefont
  {{Pashley}}},\ }\href@noop {} {\bibfield  {journal} {\bibinfo  {journal}
  {Langmuir}\ }\textbf {\bibinfo {volume} {16}},\ \bibinfo {pages} {5749}
  (\bibinfo {year} {2000})}\BibitemShut {NoStop}%
\bibitem [{\citenamefont {Moazzami-Gudarzi}\ \emph {et~al.}(2016)\citenamefont
  {Moazzami-Gudarzi}, \citenamefont {Kremer}, \citenamefont {Valmacco},
  \citenamefont {Maroni}, \citenamefont {Borkovec},\ and\ \citenamefont
  {Trefalt}}]{Moazzami-Gudarzi2016c}%
  \BibitemOpen
  \bibfield  {author} {\bibinfo {author} {\bibfnamefont {M.}~\bibnamefont
  {Moazzami-Gudarzi}}, \bibinfo {author} {\bibfnamefont {T.}~\bibnamefont
  {Kremer}}, \bibinfo {author} {\bibfnamefont {V.}~\bibnamefont {Valmacco}},
  \bibinfo {author} {\bibfnamefont {P.}~\bibnamefont {Maroni}}, \bibinfo
  {author} {\bibfnamefont {M.}~\bibnamefont {Borkovec}}, \ and\ \bibinfo
  {author} {\bibfnamefont {G.}~\bibnamefont {Trefalt}},\ }\href {\doibase
  10.1103/PhysRevLett.117.088001} {\bibfield  {journal} {\bibinfo  {journal}
  {Phys. Rev. Lett.}\ }\textbf {\bibinfo {volume} {117}},\ \bibinfo {pages}
  {088001} (\bibinfo {year} {2016})}\BibitemShut {NoStop}%
\bibitem [{\citenamefont {Moazzami-Gudarzi}\ \emph {et~al.}(2017)\citenamefont
  {Moazzami-Gudarzi}, \citenamefont {Maroni}, \citenamefont {Borkovec},\ and\
  \citenamefont {Trefalt}}]{Moazzami-Gudarzi2017}%
  \BibitemOpen
  \bibfield  {author} {\bibinfo {author} {\bibfnamefont {M.}~\bibnamefont
  {Moazzami-Gudarzi}}, \bibinfo {author} {\bibfnamefont {P.}~\bibnamefont
  {Maroni}}, \bibinfo {author} {\bibfnamefont {M.}~\bibnamefont {Borkovec}}, \
  and\ \bibinfo {author} {\bibfnamefont {G.}~\bibnamefont {Trefalt}},\ }\href
  {\doibase 10.1039/C7SM00314E} {\bibfield  {journal} {\bibinfo  {journal}
  {Soft Matter}\ }\textbf {\bibinfo {volume} {13}},\ \bibinfo {pages} {3284}
  (\bibinfo {year} {2017})}\BibitemShut {NoStop}%
\bibitem [{\citenamefont {Langmuir}(1938)}]{Langmuir1938}%
  \BibitemOpen
  \bibfield  {author} {\bibinfo {author} {\bibfnamefont {I.}~\bibnamefont
  {Langmuir}},\ }\href {\doibase 10.1063/1.1750183} {\bibfield  {journal}
  {\bibinfo  {journal} {J. Chem. Phys.}\ }\textbf {\bibinfo {volume} {6}},\
  \bibinfo {pages} {873} (\bibinfo {year} {1938})}\BibitemShut {NoStop}%
\bibitem [{\citenamefont {{Briscoe}}\ and\ \citenamefont
  {{Attard}}(2002)}]{Briscoe2002a}%
  \BibitemOpen
  \bibfield  {author} {\bibinfo {author} {\bibfnamefont {W.~H.}\ \bibnamefont
  {{Briscoe}}}\ and\ \bibinfo {author} {\bibfnamefont {P.}~\bibnamefont
  {{Attard}}},\ }\href {\doibase 10.1063/1.1500359} {\bibfield  {journal}
  {\bibinfo  {journal} {J. Chem. Phys.}\ }\textbf {\bibinfo {volume} {117}},\
  \bibinfo {pages} {5452} (\bibinfo {year} {2002})}\BibitemShut {NoStop}%
\bibitem [{\citenamefont {Trefalt}\ \emph
  {et~al.}(2017{\natexlab{b}})\citenamefont {Trefalt}, \citenamefont
  {Szilagyi}, \citenamefont {T{\'e}llez},\ and\ \citenamefont
  {Borkovec}}]{Trefalt2017}%
  \BibitemOpen
  \bibfield  {author} {\bibinfo {author} {\bibfnamefont {G.}~\bibnamefont
  {Trefalt}}, \bibinfo {author} {\bibfnamefont {I.}~\bibnamefont {Szilagyi}},
  \bibinfo {author} {\bibfnamefont {G.}~\bibnamefont {T{\'e}llez}}, \ and\
  \bibinfo {author} {\bibfnamefont {M.}~\bibnamefont {Borkovec}},\ }\href
  {\doibase 10.1021/acs.langmuir.6b04464} {\bibfield  {journal} {\bibinfo
  {journal} {Langmuir}\ }\textbf {\bibinfo {volume} {33}},\ \bibinfo {pages}
  {1695} (\bibinfo {year} {2017}{\natexlab{b}})}\BibitemShut {NoStop}%
\bibitem [{\citenamefont {{Cao}}\ \emph {et~al.}(2015)\citenamefont {{Cao}},
  \citenamefont {{Szilagyi}}, \citenamefont {{Oncsik}}, \citenamefont
  {{Borkovec}},\ and\ \citenamefont {{Trefalt}}}]{Cao2015}%
  \BibitemOpen
  \bibfield  {author} {\bibinfo {author} {\bibfnamefont {T.}~\bibnamefont
  {{Cao}}}, \bibinfo {author} {\bibfnamefont {I.}~\bibnamefont {{Szilagyi}}},
  \bibinfo {author} {\bibfnamefont {T.}~\bibnamefont {{Oncsik}}}, \bibinfo
  {author} {\bibfnamefont {M.}~\bibnamefont {{Borkovec}}}, \ and\ \bibinfo
  {author} {\bibfnamefont {G.}~\bibnamefont {{Trefalt}}},\ }\href {\doibase
  10.1021/acs.{Langmuir}.5b01649} {\bibfield  {journal} {\bibinfo  {journal}
  {Langmuir}\ }\textbf {\bibinfo {volume} {31}},\ \bibinfo {pages} {6610}
  (\bibinfo {year} {2015})}\BibitemShut {NoStop}%
\bibitem [{\citenamefont {Trefalt}(2016)}]{Trefalt2016b}%
  \BibitemOpen
  \bibfield  {author} {\bibinfo {author} {\bibfnamefont {G.}~\bibnamefont
  {Trefalt}},\ }\href {\doibase 10.1103/PhysRevE.93.032612} {\bibfield
  {journal} {\bibinfo  {journal} {Phys. Rev. E}\ }\textbf {\bibinfo {volume}
  {93}},\ \bibinfo {pages} {032612} (\bibinfo {year} {2016})}\BibitemShut
  {NoStop}%
\bibitem [{\citenamefont {Valmacco}\ \emph
  {et~al.}(2016{\natexlab{b}})\citenamefont {Valmacco}, \citenamefont
  {Elzbieciak-Wodka}, \citenamefont {Besnard}, \citenamefont {Maroni},
  \citenamefont {Trefalt},\ and\ \citenamefont {Borkovec}}]{Valmacco2016a}%
  \BibitemOpen
  \bibfield  {author} {\bibinfo {author} {\bibfnamefont {V.}~\bibnamefont
  {Valmacco}}, \bibinfo {author} {\bibfnamefont {M.}~\bibnamefont
  {Elzbieciak-Wodka}}, \bibinfo {author} {\bibfnamefont {C.}~\bibnamefont
  {Besnard}}, \bibinfo {author} {\bibfnamefont {P.}~\bibnamefont {Maroni}},
  \bibinfo {author} {\bibfnamefont {G.}~\bibnamefont {Trefalt}}, \ and\
  \bibinfo {author} {\bibfnamefont {M.}~\bibnamefont {Borkovec}},\ }\href
  {\doibase 10.1039/C6NH00070C} {\bibfield  {journal} {\bibinfo  {journal}
  {Nanoscale Horiz.}\ }\textbf {\bibinfo {volume} {1}},\ \bibinfo {pages} {325}
  (\bibinfo {year} {2016}{\natexlab{b}})}\BibitemShut {NoStop}%
\bibitem [{\citenamefont {Sader}\ \emph {et~al.}(1999)\citenamefont {Sader},
  \citenamefont {Chon},\ and\ \citenamefont {Mulvaney}}]{Sader1999}%
  \BibitemOpen
  \bibfield  {author} {\bibinfo {author} {\bibfnamefont {J.~E.}\ \bibnamefont
  {Sader}}, \bibinfo {author} {\bibfnamefont {J.~W.~M.}\ \bibnamefont {Chon}},
  \ and\ \bibinfo {author} {\bibfnamefont {P.}~\bibnamefont {Mulvaney}},\
  }\href@noop {} {\bibfield  {journal} {\bibinfo  {journal} {Rev. Sci.
  Instrum.}\ }\textbf {\bibinfo {volume} {70}},\ \bibinfo {pages} {3967}
  (\bibinfo {year} {1999})}\BibitemShut {NoStop}%
\bibitem [{\citenamefont {Hiemstra}\ \emph {et~al.}(1989)\citenamefont
  {Hiemstra}, \citenamefont {{van Riemsdijk}},\ and\ \citenamefont
  {Bolt}}]{Hiemstra1989a}%
  \BibitemOpen
  \bibfield  {author} {\bibinfo {author} {\bibfnamefont {T.}~\bibnamefont
  {Hiemstra}}, \bibinfo {author} {\bibfnamefont {W.~H.}\ \bibnamefont {{van
  Riemsdijk}}}, \ and\ \bibinfo {author} {\bibfnamefont {G.~H.}\ \bibnamefont
  {Bolt}},\ }\href@noop {} {\bibfield  {journal} {\bibinfo  {journal} {J.
  Colloid Interface Sci.}\ }\textbf {\bibinfo {volume} {133}},\ \bibinfo
  {pages} {91} (\bibinfo {year} {1989})}\BibitemShut {NoStop}%
\bibitem [{\citenamefont {Kobayashi}\ \emph {et~al.}(2005)\citenamefont
  {Kobayashi}, \citenamefont {Juillerat}, \citenamefont {Galletto},
  \citenamefont {Bowen},\ and\ \citenamefont {Borkovec}}]{Kobayashi2005}%
  \BibitemOpen
  \bibfield  {author} {\bibinfo {author} {\bibfnamefont {M.}~\bibnamefont
  {Kobayashi}}, \bibinfo {author} {\bibfnamefont {F.}~\bibnamefont
  {Juillerat}}, \bibinfo {author} {\bibfnamefont {P.}~\bibnamefont {Galletto}},
  \bibinfo {author} {\bibfnamefont {P.}~\bibnamefont {Bowen}}, \ and\ \bibinfo
  {author} {\bibfnamefont {M.}~\bibnamefont {Borkovec}},\ }\href@noop {}
  {\bibfield  {journal} {\bibinfo  {journal} {Langmuir}\ }\textbf {\bibinfo
  {volume} {21}},\ \bibinfo {pages} {5761} (\bibinfo {year}
  {2005})}\BibitemShut {NoStop}%
\bibitem [{\citenamefont {Behrens}\ and\ \citenamefont
  {Grier}(2001)}]{Behrens2001}%
  \BibitemOpen
  \bibfield  {author} {\bibinfo {author} {\bibfnamefont {S.~H.}\ \bibnamefont
  {Behrens}}\ and\ \bibinfo {author} {\bibfnamefont {D.~G.}\ \bibnamefont
  {Grier}},\ }\href@noop {} {\bibfield  {journal} {\bibinfo  {journal} {J.
  Chem. Phys.}\ }\textbf {\bibinfo {volume} {115}},\ \bibinfo {pages} {6716}
  (\bibinfo {year} {2001})}\BibitemShut {NoStop}%
\bibitem [{\citenamefont {Ackler}\ \emph {et~al.}(1996)\citenamefont {Ackler},
  \citenamefont {French},\ and\ \citenamefont {Chiang}}]{Ackler1996}%
  \BibitemOpen
  \bibfield  {author} {\bibinfo {author} {\bibfnamefont {H.~D.}\ \bibnamefont
  {Ackler}}, \bibinfo {author} {\bibfnamefont {R.~H.}\ \bibnamefont {French}},
  \ and\ \bibinfo {author} {\bibfnamefont {Y.~M.}\ \bibnamefont {Chiang}},\
  }\href@noop {} {\bibfield  {journal} {\bibinfo  {journal} {J. Colloid
  Interface Sci.}\ }\textbf {\bibinfo {volume} {179}},\ \bibinfo {pages} {460}
  (\bibinfo {year} {1996})}\BibitemShut {NoStop}%
\bibitem [{\citenamefont {Vigil}\ \emph {et~al.}(1994)\citenamefont {Vigil},
  \citenamefont {Xu}, \citenamefont {Steinberg},\ and\ \citenamefont
  {Israelachvili}}]{Vigil1994}%
  \BibitemOpen
  \bibfield  {author} {\bibinfo {author} {\bibfnamefont {G.}~\bibnamefont
  {Vigil}}, \bibinfo {author} {\bibfnamefont {Z.~H.}\ \bibnamefont {Xu}},
  \bibinfo {author} {\bibfnamefont {S.}~\bibnamefont {Steinberg}}, \ and\
  \bibinfo {author} {\bibfnamefont {J.}~\bibnamefont {Israelachvili}},\
  }\href@noop {} {\bibfield  {journal} {\bibinfo  {journal} {J. Colloid
  Interface Sci.}\ }\textbf {\bibinfo {volume} {165}},\ \bibinfo {pages} {367}
  (\bibinfo {year} {1994})}\BibitemShut {NoStop}%
\bibitem [{\citenamefont {Valle-Delgado}\ \emph {et~al.}(2005)\citenamefont
  {Valle-Delgado}, \citenamefont {Molina-Bolivar}, \citenamefont
  {Galisteo-Gonzalez}, \citenamefont {Galvez-Ruiz}, \citenamefont {Feiler},\
  and\ \citenamefont {Rutland}}]{Valle-Delgado2005}%
  \BibitemOpen
  \bibfield  {author} {\bibinfo {author} {\bibfnamefont {J.~J.}\ \bibnamefont
  {Valle-Delgado}}, \bibinfo {author} {\bibfnamefont {J.~A.}\ \bibnamefont
  {Molina-Bolivar}}, \bibinfo {author} {\bibfnamefont {F.}~\bibnamefont
  {Galisteo-Gonzalez}}, \bibinfo {author} {\bibfnamefont {M.~J.}\ \bibnamefont
  {Galvez-Ruiz}}, \bibinfo {author} {\bibfnamefont {A.}~\bibnamefont {Feiler}},
  \ and\ \bibinfo {author} {\bibfnamefont {M.~W.}\ \bibnamefont {Rutland}},\
  }\href@noop {} {\bibfield  {journal} {\bibinfo  {journal} {J. Chem. Phys.}\
  }\textbf {\bibinfo {volume} {123}},\ \bibinfo {pages} {034708} (\bibinfo
  {year} {2005})}\BibitemShut {NoStop}%
\bibitem [{\citenamefont {Grabbe}\ and\ \citenamefont
  {Horn}(1993)}]{Grabbe1993}%
  \BibitemOpen
  \bibfield  {author} {\bibinfo {author} {\bibfnamefont {A.}~\bibnamefont
  {Grabbe}}\ and\ \bibinfo {author} {\bibfnamefont {R.~G.}\ \bibnamefont
  {Horn}},\ }\href@noop {} {\bibfield  {journal} {\bibinfo  {journal} {J.
  Colloid Interface Sci.}\ }\textbf {\bibinfo {volume} {157}},\ \bibinfo
  {pages} {375} (\bibinfo {year} {1993})}\BibitemShut {NoStop}%
\bibitem [{\citenamefont {Acuna}\ and\ \citenamefont
  {Toledo}(2011)}]{Acuna2011}%
  \BibitemOpen
  \bibfield  {author} {\bibinfo {author} {\bibfnamefont {S.~M.}\ \bibnamefont
  {Acuna}}\ and\ \bibinfo {author} {\bibfnamefont {P.~G.}\ \bibnamefont
  {Toledo}},\ }\href@noop {} {\bibfield  {journal} {\bibinfo  {journal} {J.
  Colloid Interface Sci.}\ }\textbf {\bibinfo {volume} {361}},\ \bibinfo
  {pages} {397} (\bibinfo {year} {2011})}\BibitemShut {NoStop}%
\bibitem [{\citenamefont {Trompette}(2017)}]{Trompette2017}%
  \BibitemOpen
  \bibfield  {author} {\bibinfo {author} {\bibfnamefont {J.-L.}\ \bibnamefont
  {Trompette}},\ }\href@noop {} {\bibfield  {journal} {\bibinfo  {journal} {J.
  Phys. Chem. B}\ }\textbf {\bibinfo {volume} {121}},\ \bibinfo {pages} {5654}
  (\bibinfo {year} {2017})}\BibitemShut {NoStop}%
\bibitem [{\citenamefont {Trefalt}\ \emph {et~al.}(2016)\citenamefont
  {Trefalt}, \citenamefont {Behrens},\ and\ \citenamefont
  {Borkovec}}]{Trefalt2016}%
  \BibitemOpen
  \bibfield  {author} {\bibinfo {author} {\bibfnamefont {G.}~\bibnamefont
  {Trefalt}}, \bibinfo {author} {\bibfnamefont {S.~H.}\ \bibnamefont
  {Behrens}}, \ and\ \bibinfo {author} {\bibfnamefont {M.}~\bibnamefont
  {Borkovec}},\ }\href {\doibase 10.1021/acs.langmuir.5b03611} {\bibfield
  {journal} {\bibinfo  {journal} {Langmuir}\ }\textbf {\bibinfo {volume}
  {32}},\ \bibinfo {pages} {380} (\bibinfo {year} {2016})}\BibitemShut
  {NoStop}%
\bibitem [{\citenamefont {Brown}\ \emph {et~al.}(2016)\citenamefont {Brown},
  \citenamefont {Goel},\ and\ \citenamefont {Abbas}}]{Brown2016}%
  \BibitemOpen
  \bibfield  {author} {\bibinfo {author} {\bibfnamefont {M.~A.}\ \bibnamefont
  {Brown}}, \bibinfo {author} {\bibfnamefont {A.}~\bibnamefont {Goel}}, \ and\
  \bibinfo {author} {\bibfnamefont {Z.}~\bibnamefont {Abbas}},\ }\href@noop {}
  {\bibfield  {journal} {\bibinfo  {journal} {Angew. Chem., Int. Ed.}\ }\textbf
  {\bibinfo {volume} {55}},\ \bibinfo {pages} {3790} (\bibinfo {year}
  {2016})}\BibitemShut {NoStop}%
\end{thebibliography}%

\bibliographystyle{apsrev4-1}

\end{document}